\documentclass{article}

\usepackage{arxiv}

\usepackage[utf8]{inputenc} 
\usepackage[T1]{fontenc}    
\usepackage{hyperref}       
\usepackage{url}            
\usepackage{booktabs}       
\usepackage{amsfonts}       
\usepackage{nicefrac}       
\usepackage{microtype}      
\usepackage{lipsum}
\usepackage{graphicx}
 \usepackage{subfigure}
 \usepackage{float}

 \usepackage{multirow}
 \usepackage[table,xcdraw]{xcolor}

\graphicspath{ {./images/} }

\title{Comparative performance analysis of the ResNet backbones of Mask RCNN to segment the signs of COVID-19 in chest CT scans}

\author{
 Muhammad Aleem \\
  Ghulam Ishaq Khan Institute of Engineering\\
  Sciences and Technology\\
  Topi, KPK 23640 \\
  \texttt{aleemm790@gmail.com} \\
   \And
 Rahul Raj \\
  Ghulam Ishaq Khan Institute of Engineering\\
  Sciences and Technology\\
  Topi, KPK 23640 \\
  \texttt{rahule.lohana97@gmail.com} \\
  \And
 Arshad Khan, PhD \thanks{Corresponding author: \url{https://orcid.org/0000-0002-6062-0098}}\\
  Independent Researcher \\
  Machine Learning in NLP \\
  and Computer Vision, Data Science professional \\
  Southampton, UK \\
  \texttt{arshadkhanphd@gmail.com} \\
}

\begin{document}

\maketitle

\begin{abstract}

COVID-19 has been detrimental in terms of the number of fatalities and rising number of critical patients across the world. According to the UNDP (United National Development Programme) Socio-Economic programme, aimed at the COVID-19 crisis, the pandemic is far more than a health crisis: it is affecting societies and economies at their core \footnote{https://www.undp.org/content/undp/en/home/coronavirus/socio-economic-impact-of-covid-19.html}. There has been greater developments recently in the chest X-ray-based imaging technique as part of the COVID-19 diagnosis especially using Convolution Neural Networks (CNN) for recognising and classifying images. However, given the limitation of supervised labelled imaging data, the classification and predictive risk modelling of medical diagnosis tend to compromise. This paper aims to identify and monitor the effects of COVID-19 on the human lungs by employing Deep Neural Networks on axial CT (Chest Computed Tomography) scan of lungs. We have adopted Mask RCNN, with ResNet50 and ResNet101 as its backbone, to segment the regions, affected by COVID-19 coronavirus. Using the regions of human lungs, where symptoms have manifested, the model classifies condition of the patient as either "Mild" or "Alarming". Moreover, the model is deployed on the Google Cloud Platform (GCP) to simulate the online usage of the model for performance evaluation and accuracy improvement. The ResNet101 backbone model produces an F1 score of 0.85 and faster prediction scores with an average time of 9.04 seconds per inference.
\end{abstract}

\section{Introduction}

With its first emergence in Wuhan, city of China, in December 2019, the novel coronavirus, COVID-19, has engulfed the entire world rapidly, infecting more than 12 million people so far. The virus was declared as a pandemic in March 2020 by the World Health Organisation (WHO), given the fast rate at which it had been infecting people across the world. The pandemic has brought the whole world to a standstill, significantly affecting daily lives, public health, socio-economic activities and global trade. 

It is, therefore, critically important to detect those infected with the virus as early as possible so as to slow down and stop the spread of this deadly epidemic. This can only be achieved by treating the affected patients right away before they end up in the critical care of hospitals and healthcare facilities where the chances of succumbing to the virus increase manifold. Artificial Intelligence (AI) researchers, medical experts, virologists, biologists and academic researchers across the world have put their heads together to find a cure to the deadly virus. Clinical trials for various vaccines and clinical remedies are already underway but alongside all those efforts, AI has forged solutions based on medical images and CT scans. It is imperative to develop, test and deploy some supplementary diagnostic tools, based on AI algorithms, which could tackle the spread of the virus to complement the clinical and biological tool-kits. Such tools and applications will aid clinicians at the time of triage assessment, clinical investigations, and diagnosis determination \par

Recent research \cite{time_course_of_lung_changes}, employing radiology imaging techniques, for instance, suggests that such images contain salient features about the COVID-19 coronavirus. Application of the advanced AI algorithms and techniques, coupled with \textit{radiological imaging}, can play pivotal role in the accurate detection of virus-infected individuals. Moreover, AI-based technologies also have the potential of overcoming the problem of insufficient healthcare practitioners or specialist physicians operating in remote villages and socio-economically deprived areas \cite{how_5g_wireless}. 

\subsection{Are lock-downs and social distancing enough?}

COVID-19, has claimed over 625,000 \footnote{\url{https://www.worldometers.info/coronavirus/}} lives so far across the world with many more still under active treatment in hospitals, care homes and doctor surgeries. With active cases well above 15,000,000 globally, COVID-19 has spread out to 206 countries so far, prompting the respective governments all over the world to impose lock-downs, curfews and social distancing measures. So far, the only means to fight against the deadly virus has been to initially screen patients showing symptoms of COVID-19 and isolate them in order to prevent the exponential spread of the virus. The major screening method, being used worldwide, is Reverse Transcription-Polymerase Chain Reaction (RT-PCR) which detects COVID-19 from respiratory specimens. Under the current standard cycling conditions, a typical run takes approximately 20 minutes which limits the number of tests done in 24 hours time period. In addition to that, RT-PCR procedure suffers from the high false negative rates \cite{HE2020105980} \par

According to a new study \cite{paper_one}, researchers found mild to significant lung abnormalities, on the CT images of 94 percent of patients, diagnosed with COVID-19, at the time of discharge from the hospital. Such and related research studies, necessitate a follow-up monitoring of patients to prevent or slow down the spread of infection. The researchers in \cite{paper_one} performed a longitudinal study to analyze the serial CT findings over time in patients, infected with COVID-19. From January 16 to February 17, 2020, 90 patients with COVID-19 pneumonia were prospectively enrolled and followed up until they were either discharged or died until the the study was concluded. Radiologists, then, reviewed 366 CT scans for: (a) discovering potential patterns and distribution of lung abnormalities, (b) calculating the total CT scores and (c) the number of zones involved.

Following on, those features were analyzed for temporal change. The extent of CT abnormalities progressed rapidly after the onset of symptoms and peaked during the illness at day 6-11. The predominant pattern of abnormalities, after the appearance of symptoms, was Ground-Glass Opacity (GGO), where up to 62 percent CT scans of patients, showed the presence of GGOs within 5 days of symptoms' manifestation. 
Radiography assessment does produce results faster and offers a greater accessibility given that most countries have good health care system, thus, making them a good complement to RT-PCR testing. However, one of the major setbacks in developing countries is the lack of availability of expert doctors who can interpret radiography images due to the subtlety of results and indicators. To capitalise on this problem and fill up the gaps, our AI-based CT scan-based diagnostic systems can aid doctors and radiologists in interpreting the results from CT-Scans within seconds.  

This paper presents an AI-based deep learning model which outputs the location of the area, affected by the COVID-19 coronavirus and the intensity of its spread by (a) segmenting the area of lungs and (b) calculating the ratio of the area of the bounding boxes to the area of a human lung. This research also aims at reducing the prediction time of the model by harnessing the power of the Google Cloud Platform (GCP). The approach, adopted in this research, ensures that the the results are computed within a matter of seconds thus saving a good deal of doctors' and clinicians' time while dealing with COVID-19 patients. An introductory video of this research has been uploaded, as an unlisted video, on the YouTube website\footnote{\url{https://www.youtube.com/watch?v=5HNRSdJLczs&feature=youtu.be}}.

\section{Literature Review}
\label{}
Currently, the widely practised approaches to detect COVID-19 are through the use of RT-PCR, which basically detects the presence of viral nuclei acid. One of the problems with this method is the likelihood of getting a high number of false negatives. However, a recent research, published by \cite{paper_two}, reported that many patients showed the presence of GGO and collateral expansion which can be used to treat COVID-19. Out of the 1014 cases, explored by this research, chest CT scans diagnosis showed an accuracy of about 68\% and sensitivity remained 97\%. Moreover, researchers in \cite{paper_three} studied 1099 cases and reported that 86\% of the patients showed the presence of GGO, local patchy shadowing, bilateral patchy shadowing, or interstitial abnormalities. Only 2.9\% patients with severe disease did not show these symptoms. To improve upon the accuracy of chest CT diagnosis of COVID-19, several AI experts have incorporated AI into COVID-19 diagnosis due to the following reasons: (1) AI can remove human error (2) AI-based tools do not necessarily need the availability of a doctor thus addressing the issue of insufficient healthcare experts (3) AI-based approaches can aid in the post-treatment monitoring of patients which is valid, for instance, in the case of COVID-19.

Several other researchers have published their work highlighting the CT scan-based detection of the novel coronavirus. \cite{paper_four} studied 618 CT scans of which 35\% were COVID-19 patients, 36\% were Influenza-A-viral-pneumonia patients and 28\% were healthy patients. The symptoms of each disease were manually annotated by expert radiologists and the results were fed into a 3D-CNN model. This network was a simple ResNet18 network that receives the chest CT scans as input and the output was flattened into a Fully Connected Layer (FCN). Furthermore, other information, related to the location of the symptoms, were fed into the nodes and subsequently communicated to the final \textit{softmax} layer in order to predict the three target classes. The research achieved about 86.7\% of accuracy on the test samples. Another recently published study \cite{maguolo2020critic} conducted a critical evaluation of methods, used for automatically detecting COVID-19, from X-Ray images. They argue that by removing the biasness from the usual COVID-19 testing protocols, which revolves around the mere presence of COVID-19 in X-Ray images, they can improve the learning pattern of a neural network model.  

Moreover, \cite{paper_five} trained a VB-Net architecture (a custom network that combined V-Net with bottleneck structure) on a total of 249 CT scans of COVID-19 patients and used a novel Human-In-The-Loop (HITL) method to achieve higher accuracy. As part of the process, an expert radiologist identifies the symptoms himself/herself and then feeds the outcome into the VB-Net  application to output the predicted annotations. The predicted annotations are corrected by a human and re-fed into the network to achieve better predicted annotation. The system was validated on 300 CT scans of COVID-19 patients and achieved Dice similarity coefficients\footnote{The Dice similarity coefficient (DSC) was used as a statistical validation metric to evaluate the performance of both the \textit{reproducibility} of manual segmentation and the spatial overlap accuracy of automated probabilistic fractional segmentation of MR images\cite{zou2004statistical}} of 91.6\% +/- 10.0\% between automatic and manual annotations.

\cite{paper_six} went further ahead by not just taking CT scans as input but also included into the input vector the cough voice samples, fever level, nausea level, fatigue level and other symptoms. The samples were subsequently passed on to the CNN and RNN layers in order to make a prediction on the existence of COVID-19. Furthermore, the end-to-end system was integrated on a mobile platform to make it a widely accessible tool.

\section{Methodology}

This research paper draws its motivation from the current research initiatives, being carried out by various research institutions in the world relating to AI Diagnosis for COVID-19 detection and treatment with a view to integrating it on to the IoT platform. This paper presents a diagnosis that was implemented using Mask RCNN with ResNet101 and ResNet50 as backbone models and were trained on 10 and 100 epochs respectively to cross-compare the behaviour of the said models. The model outputs the location of the symptom and the intensity of the symptoms by segmenting the area of patient lungs. Furthermore, it calculates the ratio of the area of the bounding boxes and the areas of interest of the lungs in question. The best result, achieved by the model, was an accuracy of 83\% with a sensitivity of 98\% and specificity of 63\%. Moreover, the system was integrated on the GCP platform to simulate how doctors can use the system globally. The system was tested to have an average inference time of 9.04 seconds while the fastest inference was calculated in 5.36 seconds.
\label{headings}
\subsection{Data Collection and preparation}

In this research, we use a total of 669 CT scans of which 313 are of positive COVID-19 cases and 356 are patients who have either common pneumonia, lung cancer or are healthy people grouped together as non-COVID cases. The data was split, based on the \textit{70-15-15 method}, in which 70\% (n=469) CT scans are for training, 15\% (n=100) for validation and the remaining 15\% (n=100) is for testing. The data was obtained from \cite{paper_seven}, their established GitHub repository\footnote{\url{https://github.com/ieee8023/covid-chestxray-dataset}}, and Caristica Radiological COVID-19 database\footnote{\url{https://www.sirm.org/category/senza-categoria/covid-19/}} and other classified confidential sources. The final dataset was kept fairly unbiased using the distribution given in Table \ref{tab:unbiased_table}.

\begin{table}[H]
\caption{Dataset distribution}
\begin{center}
\begin{tabular}{|p {3cm} |p{3cm} |p{3cm}|}
\hline
\multicolumn{3}{|c|}{Dataset}\\
\hline
\hline
 &COVID-19&Non-COVID\\
\hline
Train&213&256\\
Validation&55&45\\
Test&45&55\\
\hline
\end{tabular}{}
\end{center}

 \label{tab:unbiased_table}

\end{table}

\begin{figure}[H]
\centering
   \subfigure[CT scan sample of a healthy patient]{
     \includegraphics[width=0.35\linewidth]{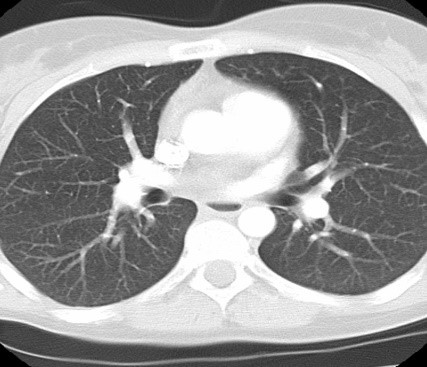}
    \label{Figure:ct_scan_healthy_pat:left}
  }
  \hspace*{1cm}
\subfigure[CT scan sample of a COVID-19 affected patient]{
   \includegraphics[width=0.35\linewidth]{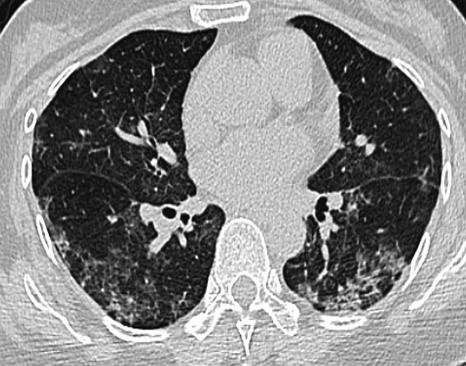}
  \label{Figure:ct_scan_COVID-19:right}
  }
  
  \caption{CT scan samples of healthy and affected patients}
\end{figure}

Moreover, Figure \ref{Figure:ct_scan_healthy_pat:left} shows a sample of non-COVID case and Figure \ref{Figure:ct_scan_COVID-19:right} shows a sample of COVID-19 case, taken from the final dataset.

\subsection{Data Pre-processing}

\begin{figure}[H]
\centering
   \subfigure[Original CT scan]{
     \includegraphics[width=0.35\linewidth]{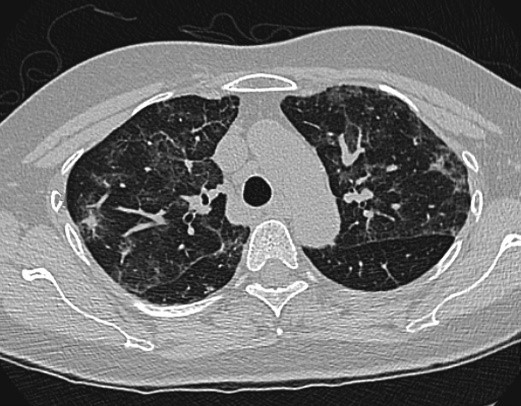}
    \label{Figure:original_ct_scn:left}
  }
  \hspace*{1cm}
\subfigure[Rotated CT scan]{
   \includegraphics[width=0.35\linewidth]{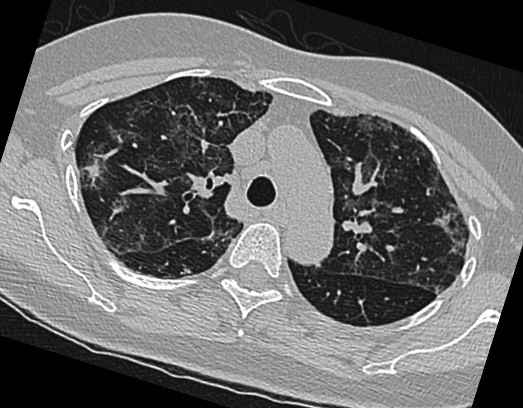}
  \label{Figure:rotated_scan:right}
  }
  
  \caption{Original vs. rotated CT scan comparison}
  \label{Figure:comparison_original_rotated}
\end{figure}

The images were first resized to (1024,1024, 3) and manually annotated using Oxford’s annotation tool\footnote{\url{http://www.robots.ox.ac.uk/~vgg/software/via/)}}. To cope up with the lack of data, augmentation techniques were applied such as rotation of 15 degrees, horizontal flipping, horizontal and vertical translations and Gaussian blur. Figure \ref{Figure:original_ct_scn:left} shows a sample image and Figure \ref{Figure:rotated_scan:right} shows the same image but in rotated form.

\subsection{Neural Network}
\begin{figure}[H]
  \centering
  \includegraphics[width=0.65\linewidth]{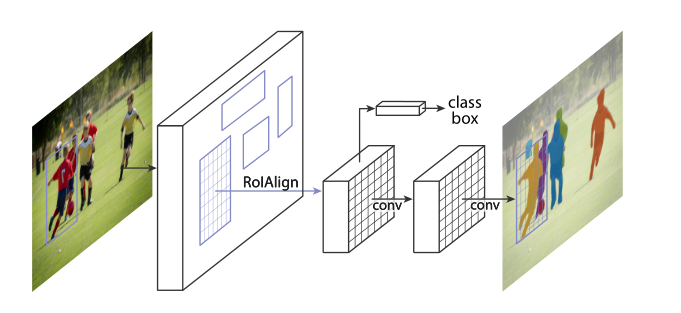}
  \caption{Network architecture of Mask RCNN} 
  \label{fig:typical_mask_r-cnn}
\end{figure}

The input images are then fed into Mask RCNN. This network uses Region Proposal Network (RPN) to generate Regions of Interest (ROIs). On the CT scans, 1000 ROIs are proposed which contain even the smallest Ground Glass Opacities. These ROIs are then fed into an ROIPool layer that extracts the features and infers the bounding box coordinates from these ROIs. The “backbone” of Mask RCNN is a neural network that is at the heart of both aforementioned processes. The backbone models, used in this research, are ResNet50 and ResNet101, which are 50 layers and 101 layers deep respectively. No pre-trained weights were used for this purpose because the CT scans are not similar to the classes on which the networks was fine-tuned. A typical Mask RCNN network, given in \cite{le2020mask} is shown in Figure \ref{fig:typical_mask_r-cnn}. 

In this research, these models were trained on 10 and 100 epochs respectively and then accuracy was evaluated using various metrics. The area of the output bounding box is then calculated, which is divided by the area of the lungs to get the Intensity of the symptoms. The intensity is then used to classify the symptoms as either “Mild” or “Alarming”. In this study, an Intensity of 0.15 was set as the threshold to classify an image as “Alarming”. Moreover, to be classified as a COVID-19 class, the model has to output at least one bounding box.

\section{Results and Discussion}

\begin{figure}[H]
   \subfigure[Training results based on ResNet50]{
    \includegraphics[width=0.55\linewidth]{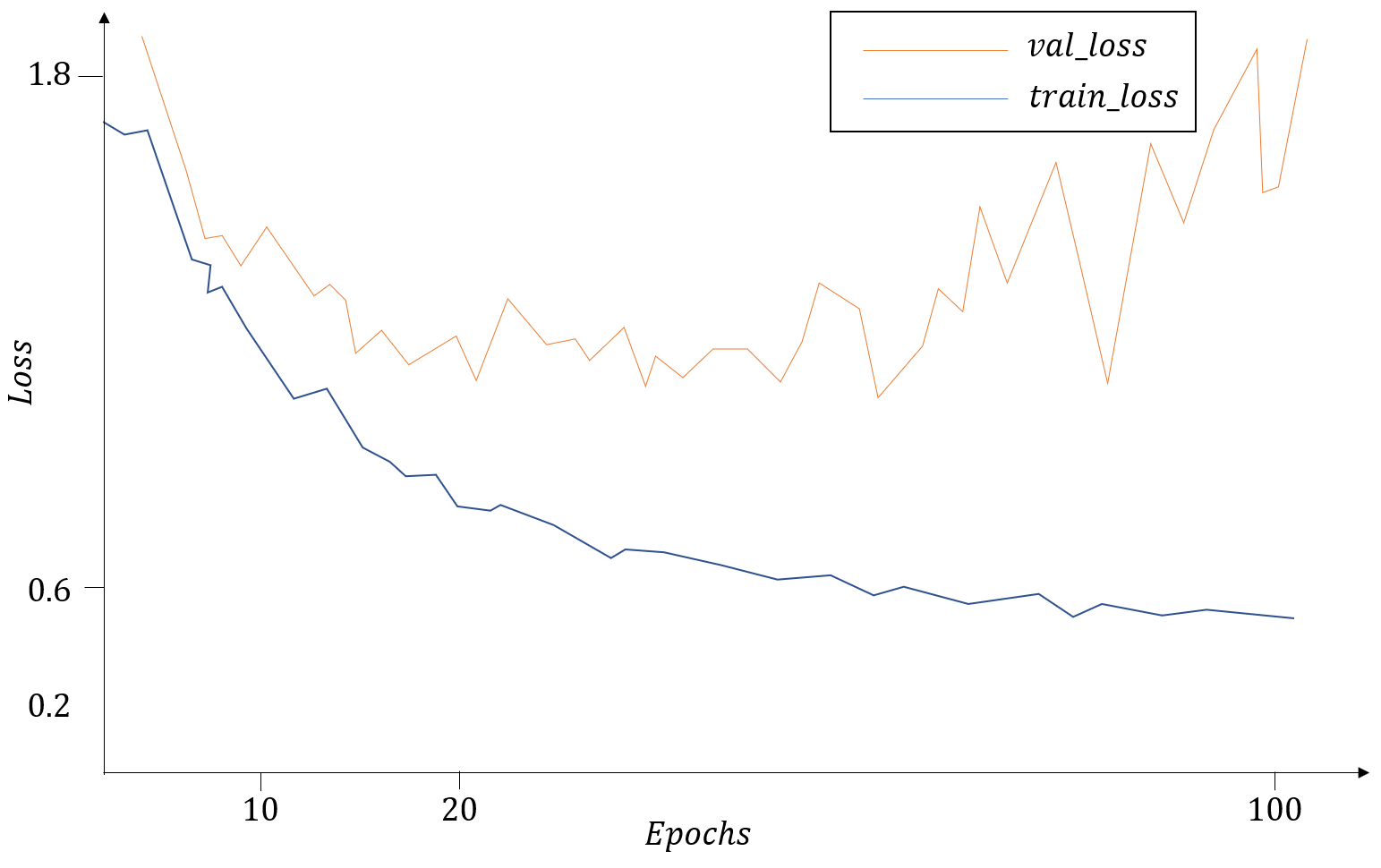}
    \label{Figure:results_resnet50:left}
  }
\subfigure[Training results based on ResNet101]{
  \includegraphics[width=0.55\linewidth]{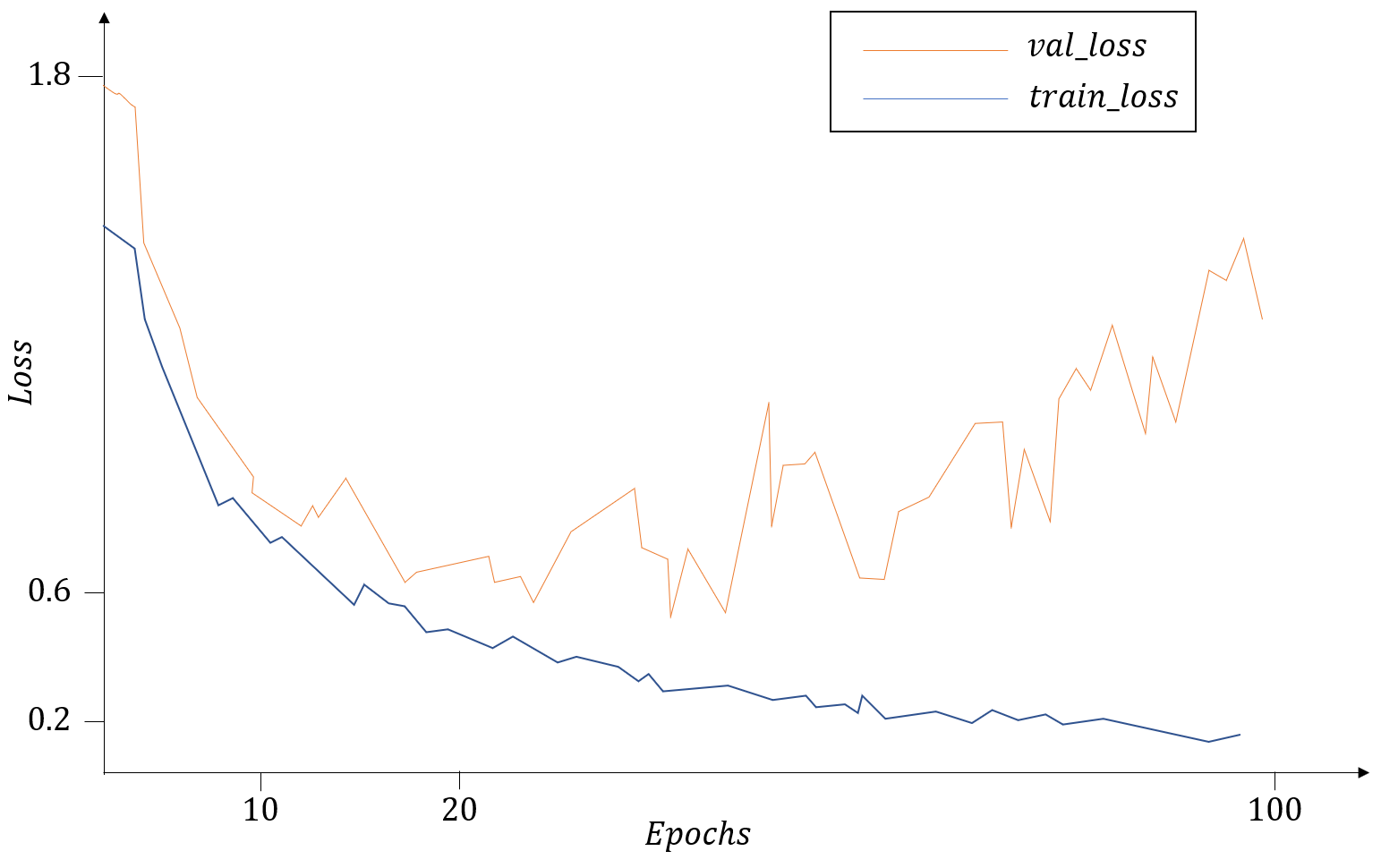}
  \label{Figure:results_resnet101:right}
  }
  
  \caption{Graphs showing training and validation losses for ResNet50 and ResNet101 from left to right}
  \label{Figure:comparison_of_resnet50_resnet101}
\end{figure}

The graphs of training loss and validation loss for ResNet50 and ResNet101 are shown in Figure \ref{Figure:results_resnet50:left} and Figure \ref{Figure:results_resnet101:right}. It can be seen that both of these models show a decreasing trend in the training loss as the number of epochs increases. However, for both models, validation loss decreased for a certain time before starting to increase after epoch 20 for both the models. This shows that the models are over-fitting. 

This can also be seen through the accuracy figures for each model. Also, the accuracy and F1 scores show that the ResNet101 is a superior model because it is 101 layers deep and able to better learn the less evident  symptoms, present in the lungs. ResNet101, trained on 10 epochs, is therefore a better choice for the purpose in question because it has produced better results; and the gap between the validation loss and training loss is minimal, meaning that the model is a good fit. The accuracy figures of both models show that the ResNet101 is better suited for this task because the complex model is comparatively better able to learn the small complex patterns of GGO in patients' lungs compared to the ResNet50. In terms of accuracy figures. ResNet101 backbone gave an accuracy of 87\%, which is the highest achieved after tuning the hyper-parameters. Compared to that, ResNet50 scored an accuracy of 73\%, which is way lower than that of ResNet101.  

We have also presented the Precision, Recall and F1-scores, produced by ResNet101 and ResNet50 backbones, in Table \ref{table:ResNet101_table_results} and Table \ref{table:ResNet50_table_results} respectively. 

\begin{table}[H]
\caption{Results produced by ResNet101}
\begin{center}
\begin{tabular}{|p {3cm} |p{3cm} |p{3cm} |p{3cm}|}
\hline
\multicolumn{4}{|c|}{ResNet101}\\
\hline
\hline
&Precision &Recall&F1-score\\
\hline
Non-COVID&0.90&0.80&0.85\\
COVID-19&0.85&0.93&0.89\\
\hline
\end{tabular}{}
\end{center}
\label{table:ResNet101_table_results}
\end{table}

\begin{table}[H]
\caption{Results produced by ResNet50}
\begin{center}
\begin{tabular}{|p {3cm} |p{3cm} |p{3cm} |p{3cm}|}
\hline
\multicolumn{4}{|c|}{ResNet50}\\
\hline
\hline
&Precision &Recall&F1-score\\
\hline
Non-COVID&0.70&0.67&0.69\\
COVID-19&0.74&0.78&0.76\\
\hline
\end{tabular}{}
\end{center}
\label{table:ResNet50_table_results}
\end{table}
As evident in the tables, ResNet101 backbone was able to achieve a 93\% recall for COVID-19 cases as compared to 80\% recall for normal cases or cases with no COVID symptoms exhibited in the CT scans of patient lungs. Whereas, for ResNet50, the recall was 78\% for COVID-19 and 67\% for normal cases. Another important variable is the F-score which helps to measure recall and precision at the same time. Since it uses harmonic mean in place of arithmetic mean, it penalises the extreme values more than the others. This research achieved an F1 score of 89\% for COVID-19 in ResNet101 and 76\% in ResNet50. Hence, it is clear that the fine-tuned Resnet101 model is better for \textit{inferencing} as compared to the ResNet50 model. The results show the efficiency of Resnet101 in detection of lesions and ground-glass opacities of infected COVID-19 patients outweighs that of ResNet50. We understand that this will aid the radiologists and clinicians at the time of testing  as well as \textit{triaging}. That in turn, will lead to taking the pressure of the respective healthcare systems and hospitals at the time of crisis. 
\subsection{Confidence intervals}
We also created a sample distribution of predictions from our model, and consequently used the distribution to calculate the confidence intervals of the original predictions. Furthermore, upper and lower limits were calculated from these predictions, for both Covid and Non-Covid cases. Table \ref{table:confidence_intervals_wilson} shows the confidence intervals estimates, using the Wilson confidence interval\footnote{\url{https://en.wikipedia.org/wiki/Binomial_proportion_confidence_interval\#Wilson_score\_interval}} formula, given in Equation \ref{eq:wilson_intervals}: 

\begin{equation}
\frac{1}{1+\frac{z^{2}}{n}}\left(\hat{p} + \frac{z^{2}}{2n}\right )\pm \frac{z}{1+\frac{z^{2}}{n}} \sqrt{\frac{\hat{p}\left ( 1-\hat{p} \right )}{{n}}+\frac{z^{2}}{4n^{2}}}
\label{eq:wilson_intervals}
\end{equation}
\textit{where} $\hat{p}$ is our metric (precision or recall), n is the sample size and z is the constant 1.96 for 95\% confidence interval.

\begin{table}[H]
\centering
\caption{\label{tab:confidence_intervals_distributions} Confidence intervals, based on predictions from ResNet backbones of MASK RCNN}

\begin{tabular}{|l|l|l|l|l|}
\hline
                   & \multicolumn{2}{c|}{\cellcolor[HTML]{C0C0C0}\textbf{Precision}} & \multicolumn{2}{c|}{\cellcolor[HTML]{C0C0C0}\textbf{Recall}} \\ \cline{2-5} 
\multirow{-2}{*}{} & \multicolumn{1}{c|}{Covid}   & \multicolumn{1}{c|}{Non-Covid}   & \multicolumn{1}{c|}{Covid}  & \multicolumn{1}{c|}{Non-Covid} \\ \hline
\textbf{ResNet50}  & 0.72+-0.13                   & 0.68+-0.12                       & 0.75+-0.13                  & 0.65+-0.12                     \\ \hline
\textbf{ResNet101} & 0.82+-0.11                   & 0.87+-0.08                       & 0.89+-0.08                  & 0.77+-0.11                     \\ \hline
\end{tabular}
\label{table:confidence_intervals_wilson}
\end{table}

\subsection{Cloud Deployment}

To simulate the environment of an online system where this model will be practically used (such as in a hospital), the ResNet101 model, trained on 10 epochs, was deployed on the GCP (Google Cloud Platform) using a Tesla K80 GPU. The user interface was designed with simplicity and practicality in mind which takes CT scans as as input before passing them on to the trained model. The data is then sent to the trained model for inference using a Flask server. The average inference time for each image was recorded as 9.04 seconds and the fastest inference was clocked at 5.36 seconds.

\subsubsection{Inference on GCP platform}
\begin{figure}[H]
  \centering
  \includegraphics[width=0.5\linewidth]{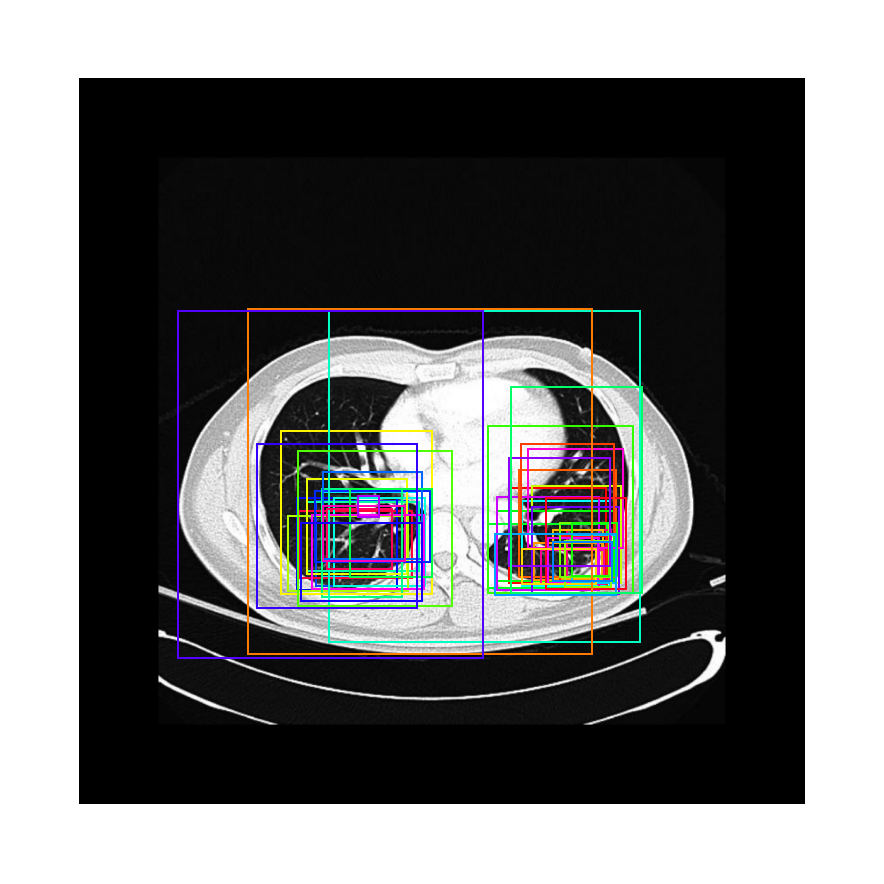}
  \caption{ROIs generated by ResNet101}
  \label{ResNet_ROIs}
\end{figure}

Figure \ref{ResNet_ROIs} shows the ROIs, proposed on a sample image of healthy lung. The ROIs have been restricted to 50 for better visualization and interpretation.

\begin{figure}[H]
  \centering
  \includegraphics[width=0.5\linewidth]{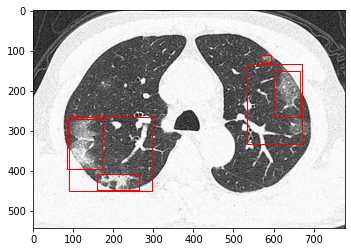}
  \caption{Bounding boxes with the highest probabilities}
  \label{probability_based_bounding_box}
\end{figure}

The ROIPool layer then selects the bounding boxes with the highest probability, as shown in Figure \ref{probability_based_bounding_box}, on a sample image of a positive COVID-19 case.

\begin{figure}[H]
  \centering
  \includegraphics[width=0.5\linewidth]{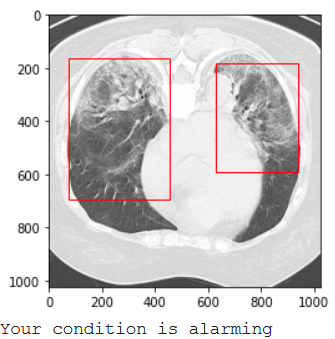}
  \caption{Bounding boxes with the highest probabilities}
  \label{highest_prob_bounding_box}
\end{figure}

Furthermore, another algorithm is implemented that removes the smaller bounding box if the intersection between the two boxes is 1. Afterwards, the ratio of the area of the bounding boxes and the area of the lungs are computed to  (a) get the intensity and (b) determine the condition of the patient, as shown in Figure \ref{highest_prob_bounding_box}.

\subsubsection{Software as a Service (SaaS)}
Using the trained model, running at the back-end, an MVP (Minimal Viable Product) was developed as a monitoring system, to enable hospitals and clinics track the COVID-19 patients. It enabled secure and faster patient management through the GCP services without having to invest in expensive software platforms. This approach also suits the IT environment in hospitals and healthcare facilities where integrating a software product into the existing infrastructure is time consuming and error-prone. 

\begin{figure}[H]
  \centering
  \includegraphics[width=0.75\linewidth]{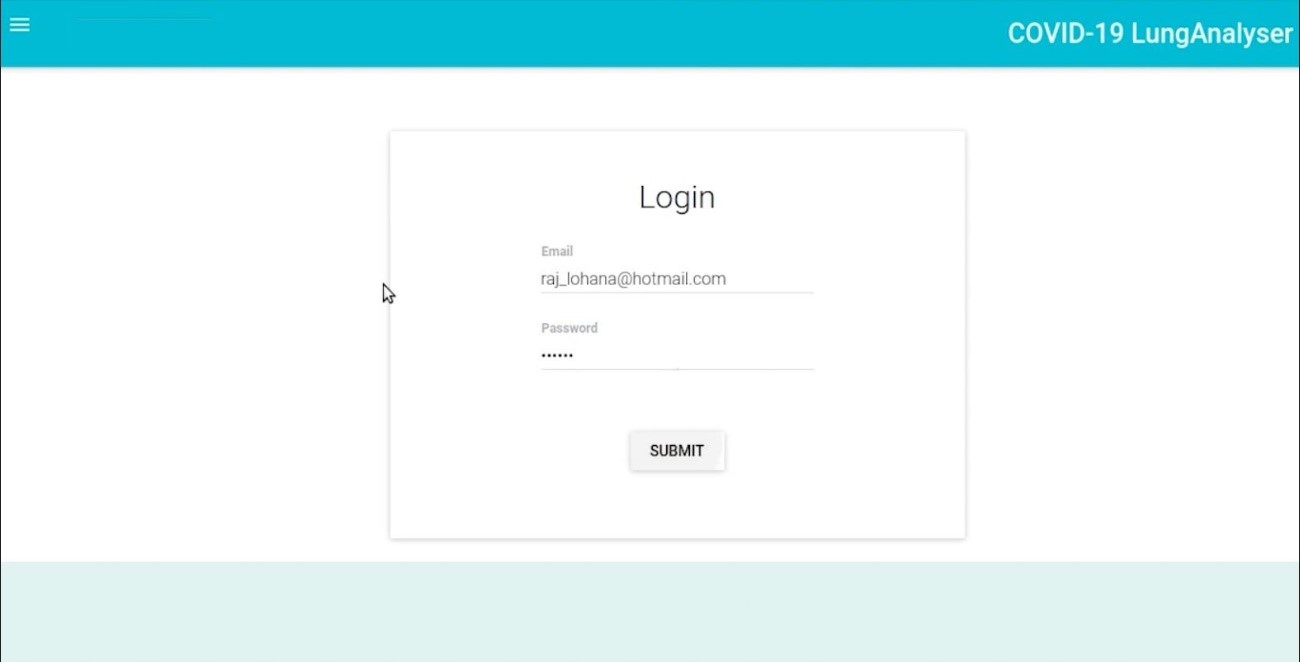}
  \caption{Login Page}
\end{figure}
The MVP service uses the trained models at the back-end to do \textit{inferencing} in the cloud and produces results on the purpose-built dashboard. Back-end for the MVP product was developed from scratch using Python programming language and the Flask framework. Endpoints for the application were made according to the RESTful routing protocol. The front-end for the MVP product was developed on React\footnote{React is an open-source JavaScript library that is used for building user interfaces specifically for single-page applications - \url{https://reactjs.org/}} and Material UI\footnote{The Material-UI Grid component is used to display the courses in a grid layout which is responsive to the screen size- \url{https://material-ui.com/}}. It allowed the clinicians and other authorised users to easily login to the application and manage their patients in real-time. PostgreSQL was used as the Relational Database Management System (RDBMS) to store all the data pertaining to the hospital, patients and the clinicians. CT Scans were stored in the Google Cloud Bucket which allowed faster processing of the previously processed data regarding each patient. JSON Web Tokens\footnote{JSON Web Token (JWT) is a toolkit for facilitating secure client-server authorization - \url{https://jwt.io/introduction/}} were used for authentication so as to make the data processing and transmission more secure. Such JWT-based authentication was also needed to allow data authentication, based on user type and access level, to ensure role-based access to the data in question.\par
The MVP was deployed using the GCP with Load Balancing enabled, which allowed for high availability requirements when the service was being used simultaneously; it ensured automatic scaling of the desired computing resources on as-needed basis.    
Since the MVP is a SaaS Platform, it allows the users to access the portal using the dedicated domain name. Doctors or other authorized users could easily login using his or her email address and password. After a successful login, the dashboard is displayed, where the user can explore features using the navigational menu to (1) access existing patients, (2) add further predictions to existing patients, (3) view the previous inferences of each patient, (4) check out the progression of the disease under the \textit{Reports tab}, which shows analytical graphs in real-time.

\begin{figure}[H]
  \centering
  \includegraphics[width=0.80\linewidth]{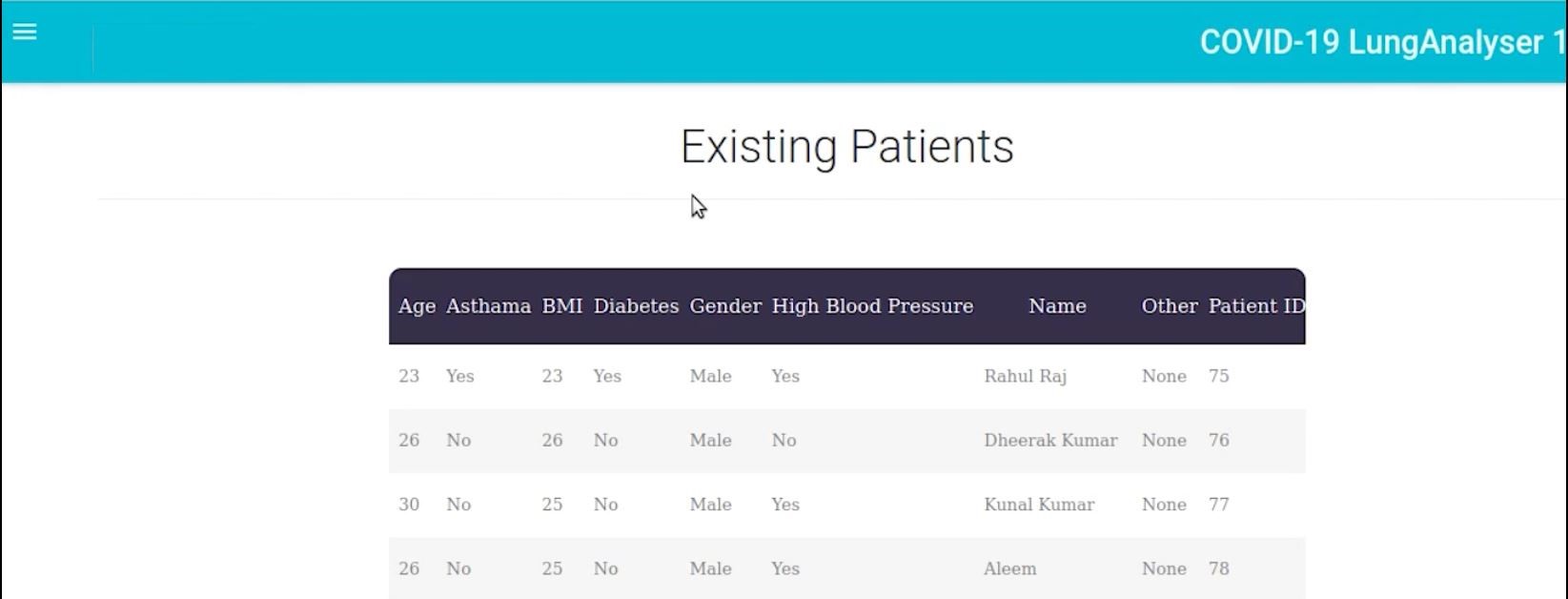}
  \caption{Patient Records of the existing patients}
  \label{patient_records}
\end{figure}

Clinicians can also add in new patients and can also explore the FAQ’s section for support and help. When a new patient option is selected, the user or clinician is prompted to add in the necessary details regarding the patient. A unique identifier is automatically created for each new user which is then stored in the PostgreSQL database at the back-end. Results, produced after the inferencing on each CT Scan of a patient, are stored inside the Google Bucket in an encrypted format. Unique URLs are then served through the cloud to the previous and current inferences (spanning over several days), which adds an additional layer of security so that only the clinicians or experts, having the desired access to the patient data, can view their respective patients' records and reports. A screen shot of the dashboard, showing patient records, can be seen in Figure \ref{patient_records}. 

\section{Conclusion}
Alternative methods of detecting COVID-19 are essential not only in the developed countries but also in developing ones where insufficient and under-resourced healthcare facilities significantly slows down the pace of testing, \textit{triaging} and treating COVID-19 patients. The application of AI-based tools and algorithms have become common place in detecting COVID-19 and the contribution, we have made through this research, is part of such efforts. We have used Mask RCNN with ResNet50 and ResNet101 backbones in order to segment the signs of COVID-19 and determine the intensity of the disease. However, we understand that the availability of more data for this research can further consolidate the performance and accuracy of our model. It is also part of further work to evaluate the model performance, in terms of better bounding box predictions, using both \textit{RoIPooling} and \textit{ROIAlign} operations with \textit{fix-sized features mapping} and \textit{bi-linear interpolation} respectively. Moreover, it is also important to highlight that the real-time feedback from clinicians in terms of refining ROIs detection in the scan will further improve the model performance at the time of training and inferencing. These kind of practices and iterative feedback mechanism can be implemented on the GCP platform subject to wider adoption of the system in healthcare facilities.   

\bibliographystyle{unsrt}

\bibliography{paper}

\end{document}